# Raman spectroscopic determination of the length, strength, compressibility, Debye temperature, elasticity, and force constant of the C-C bond in graphene

X. X. Yang,[1] J. W. Li,[1] Z. F. Zhou,[1] Y. Wang,[2] L. W. Yang,[1] W. T. Zheng,[3] Chang Q. Sun[1,4,♣]

**Abstract**

From the perspective of bond relaxation and vibration, we have reconciled the Raman shifts of graphene under the stimuli of the number-of-layer, uni-axial-strain, pressure, and temperature in terms of the response of the length and strength of the representative bond of the entire specimen to the applied stimuli. Theoretical unification of the measurements clarifies that: (i) the opposite trends of Raman shifts due to number-of-layer reduction indicate that the G-peak shift is dominated by the vibration of a pair of atoms while the D- and the 2D-peak shifts involves z-neighbor of a specific atom; (ii) the tensile strain-induced phonon softening and phonon-band splitting arise from the asymmetric response of the $C_{3v}$ bond geometry to the $C_{2v}$ uni-axial bond elongation; (iii) the thermal-softening of the phonons originates from bond expansion and weakening; and (iv) the pressure- stiffening of the phonons results from bond compression and work hardening. Reproduction of the measurements has led to quantitative information about the referential frequencies from which the Raman frequencies shift, the length, energy, force constant, Debye temperature, compressibility, elastic modulus of the C-C bond in graphene, which is of instrumental importance to the understanding of the unusual behavior of graphene.

---

[♣] Corresponding author. Tel: 65 6790 4517. Fax: +65 68733318. E-mail: ecqsun@ntu.edu.sg (Chang Q Sun)

**Table of Contents**

(For editor and referees' convenience only)

TOC art work

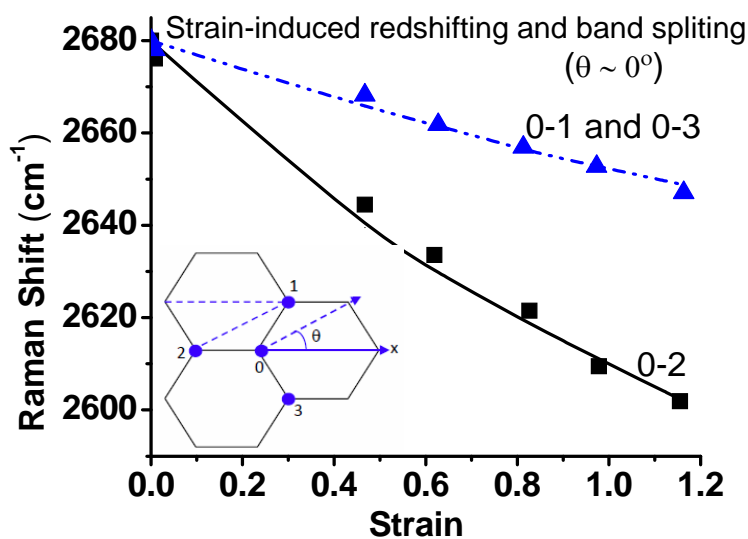

[1]Institute for Quantum Engineering and Micro-Nano Energy Technology, Key Laboratory of Low-Dimensional Materials and Application Technologies, and Faculty of Materials and Optoelectronic Physics, Xiangtan University, Hunan 411105, China

[2] School of Information and Electronic Engineering, Hunan University of Science and Technology, Xiangtan 411201, China;

[3]Department of materials Science, Jilin University, Changchun Changchun 130012, China

[4] School of Electrical and Electronic Engineering, Nanyang Technological University, Singapore 639798, Singapore

## 1. Introduction

Overwhelming contributions have been made in recent years with establishment of a huge database towards the lattice dynamics of the single- and the few-layer graphene and their nanoribbons (GNRs) under externally applied stimuli. The stimuli include the number-of-layer (n), [1, 2, 3, 4] uni-axial compressive and tensile strains (ε),[5, 6, 7] pressure (P),[8] temperature (T),[9, 10] defect density and location,[11] substrate interaction,[12, 13] hydrogenation,[14, 15] dopant,[16] incident photon polarization,[17] edge conditions,[18, 19] etc. A large graphene sheet manifests two Raman phonon modes: i) as the standard first-order approximation of Raman process, the G band (~1580 cm$^{-1}$) was suggested to arise from the in-plane vibration of the $sp^2$ carbon network;[20] ii) the 2D band (2680 cm$^{-1}$) was believed as a second-order process of double resonant Raman feature.[21] In the presence of undercoordinated defect or edge atoms, a defect-induced D band at frequencies around 1345 cm$^{-1}$ can be resolved with intensity being varied with edge conditions.[18, 19]

The Raman shifts of graphene are very sensitive to the applied stimuli. The D and 2D bands undergo a redshift when the number-of-layer of the graphene is reduced and the Raman frequencies change with the energy of the incident radiation.[3, 22] Under 514.5 nm light radiation, the D mode shifts from 1367 to 1344 cm$^{-1}$ and the 2D mode changes from 2720 to 2680 cm$^{-1}$ when the bulk graphite evolves into the monolayer graphene.[4] In contrast, a blue shift happens to the G mode that shifts from 1582 to 1587 cm$^{-1}$ when the n is reduced from 20 to one.[3, 4] When the n is increased from a few to multiple, the Raman peaks turn from the dominance of the monolayer component to the dominance of the bulk component.[3] The opposite n-dependent trends indicate that the G mode and the D/2D modes are governed by different mechanisms though their origins remain to be clear.

When the graphene is under an uni-axial tensile strain[5] or heating,[23] the Raman peaks shift to lower frequencies. When pressure is increased, the Raman frequency shifts up sublinearly.[8] The uni-axial tensile strain can soften and split the G and 2D bands and the extent of the band splitting depends on the magnitude and the relative direction of the strain to the crystal orientation.[7, 24] Under the compressive strain, the Raman shift extrapolates the trend of the tensile strain.[25] It is even amazing that the spectral intensity of the D band is one order higher at the armchair edge than that at the zigzag edge of the GNR.[18, 19]

In order to describe the effect of P and T on the Raman shifts in general, the empirically quadratic functions are often used,[26, 27]

$$\begin{cases} \omega(T) = \omega(0) + \Delta\omega_e(T) + \Delta\omega_d(T) & (\textit{Thermal softening}) \\ \omega(P) = \omega(0) + aP + bP^2 & (\textit{pressure hardening}) \end{cases}$$

where $\omega(0)$ is the frequency measured at the ambient temperature and under the atmospheric pressure. The terms of $\Delta\omega_e(T)$ and $\Delta\omega_d(T)$ represent contributions from, respectively, the thermal expansion and the coupling of anharmonic phonons of multiple branches. In the P effect, the freely-adjustable parameters, a and b, are involved. A Grüneisen parameter, $\gamma = -\partial\omega/\partial\varepsilon$ or $\gamma_E = -\partial Ln\omega/\partial Ln\varepsilon$ has also been employed to describe the strain effect.[5] Recent density-functional theory calculations[24] suggested that the strain activates the processes of "two-phonon double-resonance scattering", which are responsible for the strain-induced phonon band splitting and softening. Although the given empirical models could fit to the measurements independently, theoretically consistent formulation of the measurements remains challenging. In particular, the number-of-layer effect on the Raman shift remains theoretically unexplored.

Consistent insight into the mechanism behind the multiple-factor-stimulated Raman shift and, most strikingly, an extraction of quantitative information about the bonding identities from the sophisticated measurements are highly demanded and should be the provision of the sophisticated experiments.

A physical model should meet the following criteria and be able to:

i) reproduce the measurements with meaningful parameters;

ii) provide consistent insight into the physical mechanism behind observations;

iii) extract quantitative information from the measurements; and,

iv) establish the correlation between the measurements under various stimuli such as size, strain, pressure and temperature in the present course.

It is our thought that the external stimuli activate a certain set of intrinsic parameters that govern the Raman shift intrinsically. The vibration frequency should depend on the stimuli in a hidden form, instead. The relationship of the "vibration-intrinsic parameter-stimuli" is to be established. The objective of this contribution is to show that incorporating our original bond order-length-strength (BOLS) correlation theory[28, 29, 30, 31] to the Raman spectroscopy has enabled us to reconcile the effects of the number-of-layer, pressure, strain, and temperature on the Raman shifts of graphene. From the perspective of the local bond averaging (LBA) approach,[32] we focus on the formulation of the Raman shifts as a function of the order, length and strength of the representative bond of the entire specimen and their response to the applied stimuli. Agreement between the modeling predictions and the measurements has led to consistent insight into the mechanism behind the fascinations with revealing of quantitative information of the referential frequency $\omega(1)$ and its bulk shift, binding energy $E_b$, atomic cohesive energy $E_{coh}$, binding energy density $E_{den}$, Debye temperature $\theta_D$, elastic modulus $B$, force constant $k$, compressibility $\beta$, and the effective coordination number (CN, or *z*) for the few-layer graphene and their bond lengths and energies, which is beyond the scope of available approaches.

## 2. Principles
### 2.1 BOLS correlation

Extended from the "atomic coordination–radius" correlation premise of Goldschmidt,[33] Pauling[34] and Feibelman[35] and experimental evidence [36, 37, 38, 39, 40] to include the bond energy response to the spontaneous contraction of the bonds between undercoordinated atoms, the BOLS correlation indicates that the shorter and stronger bonds between undercoordinated atoms cause local densification and quantum entrapment of bonding electrons and binding energy, which modulate the local atomic cohesive energy, the binding energy density, the Hamiltonian of the entire specimen and the relevant properties such as mechanical strength, thermal stability, lattice dynamics, photonic, magnetic and dielectric properties associated with atomic undercoordination.[28] Numerically, the BOLS correlation is expressed as

$$\begin{cases} d_z / d_b = C_z = 2/\{1+\exp[(12-z)/8z]\} & (bond\ contraction) \\ E_z / E_b = C_z^{-m} & (bond\ strengthening) \end{cases}$$

(1)

The subscripts *z* and *b* denote an atom with z coordination neighbors and in the bulk as a standard, respectively. The bond contraction coefficient $C_z$ varies only with the effective z of the atom of concern regardless of the nature of the bond or the solid dimension. The index m = 2.56 is the bond nature indicator of carbon.[30] Using the length of 0.154 nm for the C-C bond in diamond and 0.142 nm in graphite, one can readily derive the effective CN for the bulk graphite as $z_g$ = 5.335, according to the bond contraction coefficient (eq 1). For the C atom in the bulk diamond, the effective CN is 12 instead of 4 because the diamond structure is formed by an interlock of two fcc unit cells. By the relation of $E_z = C_z^{-m}E_b$, and the known atom cohesive energy of diamond, 7.37 eV,[41] the single C-C bond energy in the diamond is $E_b$ = 7.37/12 = 0.615 eV and it is $E_3$ = 1.039 eV in the monolayer graphene of z = 3. The cohesive energy per atom in graphene is 3.11 eV/atom.

Theoretical reproduction of the elastic modulus enhancement,[30, 42, 43] melting point depression of the single-walled carbon nanotube (SWCNT),[30, 44] the C 1s binding energy shift of graphene edge, graphene, graphite, and diamond,[45, 46] the width dependence of the band gap expansion of GNR[47] and the Dirac-Fermi polarons generation and hydrogenation[29] have confirmed consistently that the C-C bond at the graphene edge contracts by 30% from 0.154 to 0.107 nm with a 152% bond energy gain.[30, 42, 43] For the 3-coordinated GNR interior atoms, the C-C bond contracts by 18.5% to 0.125 nm with a 68% increase of bond energy.[42] The Young's modulus of the SWCNT was determined to be 2.595 TPa with respect to the bulk modulus of 865 GPa. The effective wall thickness of the SWCNT is determined to be 0.142 nm. The CN reduction-induced bond contraction is in good accordance with the supershort Cr-Cr dimer distance of 0.180 nm compared with that in the bulk of 0.254 ~ 0.270 nm.[48, 49] These findings agree well with what discovered by Girit et al[50] in their transmission electron microscopic study of the thermodynamic behavior of graphene. They found that breaking a C-C bond of the 2-coordinated carbon atom near the vacancy requires 7.50 eV per bond that is 32% higher than the energy (5.67 eV/bond) required for breaking one bond of a 3-coordinated carbon atom in a suspended graphene. These findings provide further evidence for the BOLS prediction of the shorter and stronger bonds between undercoordinated carbon atoms.

## 2.2 The Raman shifts

It is emphasized that the solution to the Hamiltonian of a vibration system is a Fourier series with multiple terms of frequencies being fold of that of the primary mode.[51] Therefore, the frequency of the secondary 2D mode should be twofold that of the primary D mode. This fact may clarify the origin of the 2D mode as commonly referred as the double resonant Raman process. Any perturbation to the Hamiltonian such as the interlayer Van der Waals force or the dipole-dipole interaction or the nonlinear effect may cause the folded frequencies to deviate from the ideal values. The fact that the number-of-layer reduction induced D peak shifting from 1367 to 1344 cm$^{-1}$ and the 2D peak shifting from 2720 to 2680 cm$^{-1}$, is right within this expectation. The opposite trends of the Raman shifts due to the change of the number-of -layer indicate that the G mode is different from the D/2D mode in origin; therefore, one cannot expect to unify them simultaneously using an identical model. On the other hand, the applied strain, pressure, temperature or the atomic-CN variation can modulate the length and energy of the involved bonds, or their representative, and hence the phonon frequencies in terms of bond relaxation and vibration. Band splitting is expected to happen if the uni-axial strain is applied mismatching the graphene of $C_{3v}$ group symmetry. The extent of band splitting under strain depends on the extent of the mismatch between crystal geometry and strain.

## 2.3 Analytical solutions

Generally, one can measure the Raman resonance frequency as $\omega_x = \omega_{x0} + \Delta\omega_x$ (x = D, 2D, G), where $\omega_{x0}$ is the reference point from which the Raman shift $\Delta\omega_x$ proceeds under the applied stimuli. The $\omega_{x0}$ may vary with the frequency of the incident radiation and substrate conditions but not the nature and the trends induced by the applied stimuli. By expanding the interatomic potential in a Taylor series around its equilibrium and considering the effective atomic z, we can derive the vibration frequency shift of the harmonic system,

$$u(r) = \sum_{n=0} \left( \frac{d^n u(r)}{n! dr^n} \right)_{r=d_z} (r-d_z)^n$$

$$\cong E_z + 0 + \left. \frac{d^2 u(r)}{2! dr^2} \right|_{r=d_z} (r-d_z)^2 + 0\left( (r-d_z)^{n\geq 3} \right)..$$

$$= E_z + \frac{\mu \omega^2 (r-d_z)^2}{2} + 0\left( (r-d_z)^{n\geq 3} \right).....$$

From the dimensionality analysis, the term $\left. \frac{\partial u(r)}{\partial r^2} \right|_{r=d}$ is proportional to $\frac{E_z}{d^2}$. Equaling the vibration energy to the third term in the Taylor series and omitting the higher order terms, we have,

$$\frac{1}{2}\mu(\Delta\omega)^2 x^2 \cong \frac{1}{2} \left. \frac{\partial u(r)}{\partial r^2} \right|_{r=d} x^2 \propto \frac{1}{2}\frac{E_z}{d^2} x^2$$

As the first-order approximation, the lattice vibration frequency $\omega$ can be detected as Raman shift $\Delta\omega_x(z, d_z, E_z, \mu)$ from the reference point, $\omega_x(1, d_b, E_b, \mu)$, which depends functionally on the order z, length $d_z$, and energy $E_z$ of the representative bond for the entire specimen and the reduced mass of the dimer atoms of the representative bond with $\mu = m_1 m_2 / (m_1 + m_2)$,

$$\Delta\omega_x(z, d_z, E_z, \mu) = \omega_x(z, d_z, E_z, \mu) - \omega_x(1, d_b, E_b, \mu)$$

$$= \omega = \sqrt{\left. \frac{d^2 u(r)}{\mu dr^2} \right|_{r=d_z}} \propto \frac{1}{d_z}\left(\frac{E_z}{\mu}\right)^{1/2} \times \begin{cases} 1 & \text{(G mode)} \\ z & \text{(D/2D mode)} \end{cases}$$

(2)

The number-of-layer (coordination number) reduction induced D/2D mode redshift and the G mode blue shift suggest that the G mode be dominated by interaction between two neighboring atoms while, as the double phonon resonance, the D/2D modes involve all the z neighbors of a specific atom. According to the BOLS, the $E_z/d_z$ increases when the z is reduced, and hence blueshift happens, which is the case of the G mode; however, if the z is involved, the $zE_z/d_z$ drops with z, which is right the case of the D and 2D mode under the reduction of the number-of-layer. This clarification may deepens our insight into the bond origin of the Raman active mode not only in graphene but also in $TiO_2$ [52] with the size induced abnormal blueshift of the 141 cm$^{-1}$ vibration.

The Raman shift $\Delta\omega_x(z,d_z,E_z,\mu)$ is a hidden function of the stimuli of T and P. For convenience, we may rename $\omega_{x0}$ as $\omega_x(1)$ that is to be obtained from matching theory to the measurements. The reduced mass of the dimer $\mu$ remains a constant unless chemical adsorption or isotope is being involved. Therefore, the Raman shift provides a powerful tool for detecting any change of the order, length, strength and the reduced mass of the representative bond for the specimen. Here we use the proportional relations based on the dimensionality analysis as we do not need the exact values of the hided constants in the numerical expressions in seeking for relative change of the concerned quantities. What we are concerned are the bonding origins of the Raman shifts and the quantitative information extracted from the sophisticated measurements.

Incorporating the variables of atomic coordination, strain, temperature, and pressure ($z$, $\varepsilon$, T, P) into the expressions (2), we have the general form of the relative Raman shift,

$$\frac{\omega(z,\varepsilon,P,T)-\omega(1,\varepsilon,P_0,T_0)}{\omega(z_b,0,P_0,T_0)-\omega(1,0,P_0,T_0)} = \frac{zd_b}{d(z,\varepsilon,P,T)}\left(\frac{E(z,\varepsilon,P,T)}{E_b}\right)^{1/2}$$

where

$$\begin{cases} d_b = d(z_b,0,P_0,T_0) \\ E_b = E(z_b,0,P_0,T_0) \end{cases}$$

$$\begin{cases} d(z,\varepsilon,P,T) = d_0 \Pi_J (1+\varepsilon_J) = d_b \left[ (1+(C_z-1))\left(1+\int_0^\varepsilon d\varepsilon\right)\left(1+\int_{T_0}^T \alpha(t)dt\right)\left(1-\int_{P_0}^P \beta(p)dp\right) \right] \\ E(z,\varepsilon,P,T) = E_0\left(1+\sum_J \Delta_J\right) = E_b\left[ 1 + \dfrac{(C_z^{-m}-1) - d_z^2 \int_0^\varepsilon \kappa(\varepsilon)\varepsilon d\varepsilon - \int_{T_0}^T \eta(t)dt - \int_{V_0}^V p(v)dv}{E_b} \right] \end{cases}$$

(3)

$T_0$ and $P_0$ are the references at the ambient conditions. $\Delta_J$ is the energy perturbation and $\varepsilon_J$ the strain caused by the applied stimuli. The summation and the production are preceded over all the $J$ stimuli of $(z,\varepsilon,P,T)$. The $\alpha(t)$ and $\beta(p)$ are, respectively, the thermal expansion coefficient and the compressibility. The $k(\varepsilon)$ is the effective force constant and $\eta(t)$ the specific heat of the representative bond. These expressions indicate that the mechanical work hardening by compression or by the compressive strain will shorten and strengthen but the thermal vibration or the tensile strain will elongate and weaken the C-C bond. Atomic CN reduction shortens and strengthens the C-C bond, according to the BOLS correlation. The generalized form indicates that we can consider all the stimuli either individually or collectively, depending on the experimental conditions.

## 2.4 Number-of-layer and strain dependence

Taking $z_g = 5.335$ for the bulk graphite as a reference, we can derive from eqs (2) and (3) the reference of $\omega_x(1)$ and the z-dependent frequency $\omega_x(z)$ for the possible Raman modes. Letting

$$C_x(z,z_g) = \frac{\omega_x(z) - \omega_x(1)}{\omega_x(z_g) - \omega_x(1)} = \left(\frac{C_z}{C_{zg}}\right)^{-(m/2+1)} \times \begin{cases} \dfrac{z}{z_g} & \text{(D and 2D)} \\ 1 & \text{(G mode)} \end{cases},$$

We have,

$$\begin{cases} \omega_x(1) = \dfrac{\omega_x(z) - \omega_x(z_g)C_x(z,z_g)}{1 - C_x(z,z_g)} \\ \omega_x(z) = \omega_x(1) + [\omega_x(z_g) - \omega_x(1)]C_x(z,z_g) \end{cases}$$

(4)

From the matching to the number-of-layer dependence, we can derive the $\omega_x(1)$, the $\omega_x(z)$, and the correlation between the n and the effective z of the n-layered graphene.

Similarly, the strain-effect is given as,

$$\frac{\omega_x(z,\varepsilon) - \omega_x(1,0)}{\omega_x(z,0) - \omega_x(1,0)} = \frac{d_z}{d(z,\varepsilon)}\left(\frac{E(z,\varepsilon)}{E_z}\right)^{1/2} = \frac{\left(1 - d_z^2\int_0^\varepsilon \kappa\varepsilon d\varepsilon / E_z\right)^{1/2}}{1+\varepsilon}$$

$$\cong \frac{\left(1 - \kappa'(\lambda\varepsilon')^2\right)^{1/2}}{1 + \lambda\varepsilon'}$$

with $\kappa' = \kappa d_z^2/(2E_z) = const.$

(5)

In order to reflect the asymmetric responses of the $C_{3v}$ bond geometry to the applied $C_{2v}$ strain, we introduced a strain coefficient $\lambda$ bounded by $0 \leq \lambda \leq 1$. The $\varepsilon = \lambda\varepsilon'$ is for the branch responding less to the applied strain. $\kappa = 2E_z\kappa'd_z^{-2}$ is the effective force constant of all the $C_{3v}$ bonds of a given atom. Practically, one can measure the average strain of the entire specimen other than that of the individual bond with a force constant $\kappa_0$. Because of the $C_{3v}$ symmetry of the graphene, the applied uni-axial or bi-axial strain differentiates the actual strains of the three bonds in the graphene. The magnitudes of the strain and the strain energy of each bond should vary with the relative direction between the strain and the crystal orientation. It is therefore not surprising that the mechanical strain induces phonon band splitting.

By definition, we can derive the Grüneisen parameter from (5),

$$\gamma_E = -\frac{\partial Ln\omega'}{\partial \ln \varepsilon} = \varepsilon \left[ \frac{1+\kappa'\varepsilon}{(1-\kappa'\varepsilon^2)(1+\varepsilon)} \right]$$

$$or, \gamma'_E = -\frac{\partial \omega'}{\partial \varepsilon} = \frac{1+\kappa'\varepsilon}{(1-\kappa'\varepsilon^2)^{1/2}(1+\varepsilon)^2}$$

$$\text{With } \omega' = \frac{\omega_x(z,\varepsilon) - \omega_x(1,0)}{\omega_x(z,0) - \omega_x(1,0)}$$

(6)

From matching the strain dependent, we can derive the force constant of the C-C bond and the strain coefficient λ, without needing the Grüneisen parameter that is not a constant as a matter of fact.

## 2.5 Pressure and temperature dependence

Likewise, using the approximation $1 + x \cong \exp(x)$ at $x \ll 1$, we can formulate the thermal and pressure effects,[53]

$$\frac{\omega(z,y) - \omega(1,y_0)}{\omega(z,y_0) - \omega(1,y_0)} \cong \begin{cases} (1-\Delta_T)^{1/2} \exp\left(\int_{T_0}^T \alpha dt\right) \\ (1+\Delta_P)^{1/2} \exp\left(\int_{P_0}^P \beta dp\right) \end{cases} \quad (y = T, P)$$

(7)

The thermally- and mechanically-induced energy perturbations $\Delta_T$ and $\Delta_P$ follow the relations, [32]

$$\Delta_T = \int_{T_0}^T \frac{\eta(t)dt}{E_z} = \int_0^T \frac{C_V(T/\theta_D)}{zE_z}dt = \int_0^T \frac{4R(T/\theta_D)^3 \int_0^{\theta_D/T}(e^x-1)^{-1}x^2 dx}{E_{coh}}dt$$

$$\Delta_P = -\int_{V_0}^V \frac{p(v)dv}{E_z} = -\frac{V_0}{E_z}\int_1^x p(x)dx = -\int_1^x \frac{p(x)dx}{E_{den}}$$

$$\text{with } \begin{cases} p(x) = 3/2 \times B_0 (x^{-7/3} - x^{-5/3})(1+3(B_0'-4)(x^{-2/3}-1)/4) & (BM) \\ x(P) = V/V_0 = 1 - \beta P + \beta' P^2 & (Non-linear) \end{cases}$$

$$\tag{8}$$

The $\Delta_T$ is the integral of the specific heat reduced by the bond energy in two-dimensional Debye approximation. When the measuring temperature T is higher than $\theta_D$, the two-dimensional specific heat $C_v$ approaches a constant of 2R (R is the idea gas constant). The atomic cohesive energy $E_{coh} = zE_z$ and the $\theta_D$ are the uniquely adjustable parameters in calculating the $\Delta_T$. The $\Delta_P$ is calculated based on the integral of the Birch-Mürnaghan (BM) equation,[54, 55] or the x(p). $V_0$ is the initial volume of a bond. The variables in $\Delta_P$ are the binding energy density $E_{den} = E_z/V_0$ and the compressibility β and β'. The x(P) is another form of the equation of states in terms of the nonlinear compressibility. Matching the BM equation to the x(P) or the measured x-P curve, one can derive the nonlinear compressibility β and β', the bulk modulus $B_0$ ($\beta B_0 \cong 1$) and its first-order differentiation $B'_0$. Substituting the integrals (8) into (7), we can reproduce the P- and T-dependent Raman shift with derivatives of the $\theta_D$, α, and $E_{den}$ and the compressibility derived from the x(P) relation with the known $E_{coh}$.

3. **Experimental and calculation procedures**

The well established Raman database as functions of the number-of-layer,[3, 22] compressive [5, 25] and tensile[25] strain, temperature,[56] and pressure [8] enabled the verification of the derived formulations and expectations. In numerical calculations, the known effective bond length $d_g$ = 0.142 nm and $z_g$ = 5.335 for the bulk graphite, z = 3 for the single layer graphene, and the m = 2.56 for carbon were taken as input parameters. We assumed the experimental database is sufficiently accurate. Errors in measurements will render the accuracy of the derivatives but not the nature and the trends of the observations.

We firstly calculated the ω-z curves and calibrate them with the known vibration frequencies for the bulk graphite and the monolayer graphene. This calibration leads to the quantification of $\omega_x(1)$. Based on the z-dependent ω-z curve, we can determine the correlation between the effective atomic CN and the number-of-layer n for the few-layer graphene of all the possible modes. Likewise, matching to the strain-dependent Raman shifts and the strain-induced band splitting, we can determine the force constant and the effective strain for different branches of the splitting band. Matching to the T- and P-dependence, we can obtain the $\theta_D$, $E_{coh}$, $E_{den}$ and β and β′ values without involving any other adjustable parameters. We have thus unified, for the first time, the Raman shifts of various modes under the considered stimuli, which have not been realized by any other approaches.

## 4. Results and discussion

### 4.1 Number-of-layer dependence

With the given Raman frequencies of the 2D peak shifting from 2720 to 2680 cm$^{-1}$ and the D peak from 1367 to 1344 cm$^{-1}$ when the graphite ($z_g$) turns to be the monolayer ($z$ = 3) graphene,[3, 22, 57] and the G mode shifting from 1582 to 1587 cm$^{-1}$,[3, 4] we can calibrate the z-dependent relative shift of the possible modes,

$$C_x(z, z_g) = \frac{\omega_x(z) - \omega_x(1)}{\omega_x(z_g) - \omega_x(1)} = \left(\frac{C_z}{C_{z_g}}\right)^{-2.28} \begin{cases} \frac{z}{z_g} & (D, 2D) \\ 1 & (G) \end{cases} \text{ and,}$$

$$C_x(3, z_g) = \left(\frac{0.8147}{0.9220}\right)^{-2.28} \times \begin{cases} \frac{3.0}{5.335} & = 0.7458 \, (D, 2D) \\ 1 & = 1.3260 \, (G) \end{cases}.$$

The reference frequencies and the general expression for the z-dependency are derived as,

$$\omega_x(1) = \frac{\omega_x(3) - \omega_x(z_g) C_x(3, z_g)}{1 - C_x(3, z_g)} = \begin{cases} 1276.8 & (D) \\ 2562.6 & (2D) \\ 1566.7 & (G) \end{cases} (cm^{-1})$$

$$\omega_x(z) = \omega_x(1) + [\omega_x(z_g) - \omega_x(1)]C_x(z, z_g)$$
$$= \begin{cases} 1276.8 + 90.2 \times C_D(z, z_b) & (D) \\ 2562.6 + 157.4 \times C_D(z, z_b) & (2D) \\ 1566.7 + 16.0 \times C_G(z, z_b) & (G) \end{cases} (cm^{-1})$$

(9)

Figure 1 shows the modeling reproduction of the z dependence Raman frequencies of (a) the D/2D modes [3, 5, 25] and (b) the G mode. The inset in (b) shows the original experimental ω-n data for the G mode.[3, 4] Panel (c) is the z - n transformation derived from (a) and (b). It is seen that when the n is greater than 6, the z reaches and then maintains almost the bulk graphite value of 5.335. The consistency between predictions and the measurements of the z-dependent Raman shifts and the z-n transformation function for the three modes evidences the essentiality and appropriateness of the proposed mechanisms for the lattice vibration in graphene.

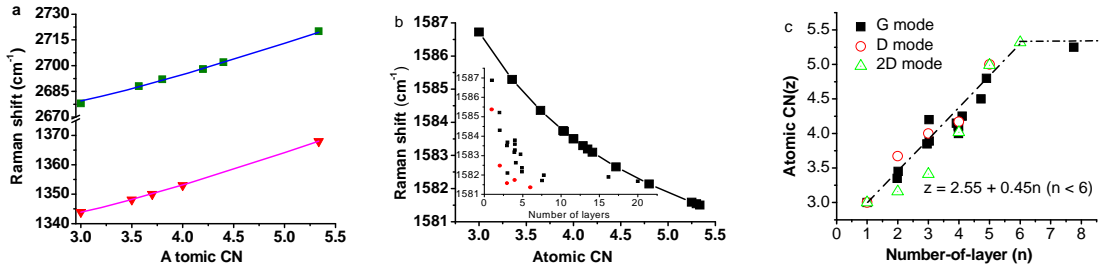

Figure 1 BOLS reproduction of the z dependence of the (a) D/2D modes [3, 22] and the (b) G mode Raman frequency with derivative of the referential vibration frequency of each. The inset in (b) shows the originally experimental ω-n data for the G mode.[3, 4] Panel (c) shows the correlation function between the atomic CN and the number-of-layer. For n > 6, the z approaches and maintains almost the bulk graphite value of 5.335.

## 4.2 Strain-induced red shifting and band splitting

Combining eq (5) and (9), we have the joint $z$- and $\varepsilon$-dependent Raman shifts:

$$\omega_x(z,\varepsilon) = \omega_x(1,0) + [\omega_x(z_g,0) - \omega_x(1,0)]C_x(z,z_g) \times (1 - \kappa'(\lambda\varepsilon)^2)^{1/2} \times (1 + \lambda\varepsilon)^{-1}$$

$$= \begin{Bmatrix} 1276.8 \\ 2562.6 \\ 1566.7 \end{Bmatrix} + C_x(z, z_g) \frac{(1 - \kappa'(\lambda\varepsilon)^2)^{1/2}}{1 + \lambda\varepsilon} \times \begin{cases} 90.2 & (D) \\ 157.4 & (2D) \\ 16.0 & (G) \end{cases} (cm^{-1})$$

(10)

Figure 2 shows the (a) uni-axial compressive [5, 25] and tensile[25] strain induced D/2D mode red shifting without band splitting and the (b) tensile strain induced 2D mode red shifting and band splitting.[24] For all the D/2D and G modes, the reduced force constant $\kappa' = \kappa d_z^2/(2E_z) = 0.30$, has been derived, corresponding to $\kappa = 6.283$ N/m for graphene ($z = 3$). The inset in (a) illustrates the asymmetric response of the three bonds of $C_{3v}$ group symmetry, denoted 1, 2, and 3, to the uni-axial tensile strain. There are two extreme situations: at $\theta = 0°$, or the strain is along bond 2, $\varepsilon_1 = \varepsilon_3 = \lambda\varepsilon_2 < \varepsilon_2$; at $\theta = 30°$, or the strain is perpendicular to bond 3, $\varepsilon_1 = \varepsilon_2 > \varepsilon_3 \sim 0$. The $\varepsilon_2$ seems to be the maximum at $\theta = 0°$. The bond configuration symmetry allows us to focus on the angle ranging from 0° to 30° between a specific bond and the strain. There should be a branch remaining the original frequency as $\varepsilon_3 \sim 0$ at $\theta = 30°$ in panel (a) ; panel (b) should correspond to $\theta = 0°$ with band splitting.

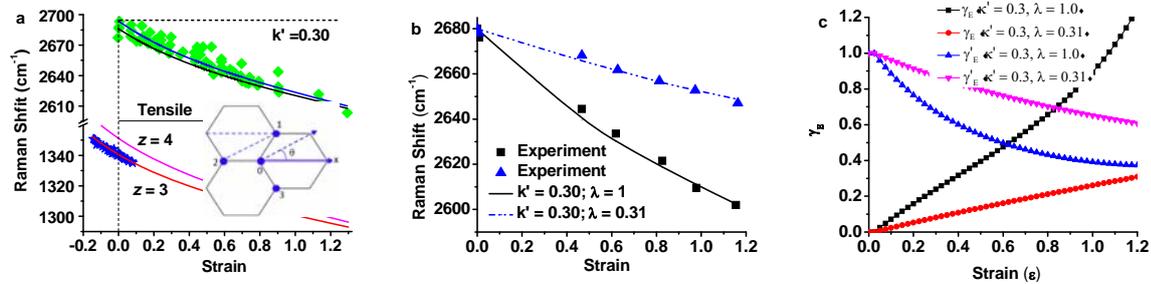

Figure 2 BOLS reproduction of (a) the uni-axial compressive[5, 25] and tensile[25] strain induced Raman shift of the D/2D bands and (b) the tensile strain induced 2D band red shifting and band splitting.[24] The inset in (a) illustrates the anisotropic response of the $C_{3v}$ bonds to the uni-axial strain. There are two extreme situations: at $\theta = 0°$, $\varepsilon_1 = \varepsilon_3 = \lambda\varepsilon_2 < \varepsilon_2$; at $\theta = 30°$, $\varepsilon_1 = \varepsilon_2 > \varepsilon_3 \sim 0$. The reduced force constant $\kappa' = 0.3$ and the effective strain of the slow-shifted branch in (b) $\varepsilon' = \lambda\varepsilon$. There should be a constant branch in (a) if $\theta = 30°$. The 2D mode in panel (a) appears to correspond to $\theta = 30°$ needing a constant branch because $\varepsilon_3 \sim 0$ and (b) to $0°$ situation. Panel (c) and eq (6) show the Grüneisen parameters for the diverged branches of the 2D mode in (b).

In addition to the Grüneisen parameters derived in eq (6), we employed the strain coefficient $\lambda$ for the strain-induced red shifting and band splitting. The introduction of the $\lambda$ and the $\kappa$ is more convenient and physically meaningful than using the Grüneisen parameter alone. In panel (b), the $\lambda = 0.31$ for the upper and $\lambda = 1.0$ for the lower branches. This result means that the bonds labeled 1 and 3 in the inset are elongated by 31% of that of the bond 2 if the stain is along bond 2. As the $\lambda$ changes with the relative direction between the strain and the crystal orientation, any possible extent of splitting and frequency variation with strain can be reproduced, which could further explain how the "two-phonon double-resonance" proceeds under the given circumstances.

From the $C_{3v}$ bond configuration shown as inset in Figure 2a, and the derived effective force constant $\kappa = 6.283$ N/m, we can estimate the force constant of the C-C bond in the single layer graphene. We may define the C-C bond force constant $k_0$, the bonds labeled 1 and 3 are approximated as in parallel and the resultant $\kappa_{13} = 2\kappa_0$. This resultant bond connects with bond 2 in series and therefore the resultant force constant of the three bonds is $\kappa_{123} = \kappa = 2\kappa_0/3$. Hence, the C-C bond force constant $\kappa_0 = 3\kappa_0/2 \sim 9.424$ N/m.

### 4.3 Pressure and temperature dependence

The matching to the measured T-dependent Raman shift of the 2D mode in Figure 3a turns out that $\theta_D = 540$ K, with the given atomic cohesive energy of 3.11 eV/atom. The $\theta_D$ is about 1/3 fold of the melting point 1605 K for the SWCNT.[30] At $T \sim \theta_D/3$, the Raman shift turns gradually from the nonlinear to the linear form when the temperature is increased. The slow decrease of the Raman shift at very low temperatures arises from the small $\int_0^T \eta dt$ values as the specific heat $\eta(t)$ is proportional to $T^2$ for the two-dimensional system at very low temperatures. These results imply that the $\theta_D$ determines the width of the shoulder and the $1/E_{coh}$ and the thermal expansion coefficient determine the slope of the curve at high temperatures. The match to the measured P-dependent Raman shift in Figure 3b gives rise to the compressibility of $\beta = 1.145 \times 10^{-3}$ (GPa$^{-1}$) and $\beta' = 7.63 \times 10^{-5}$ (GPa$^{-2}$) and the energy density of 320 eV/nm$^3$.

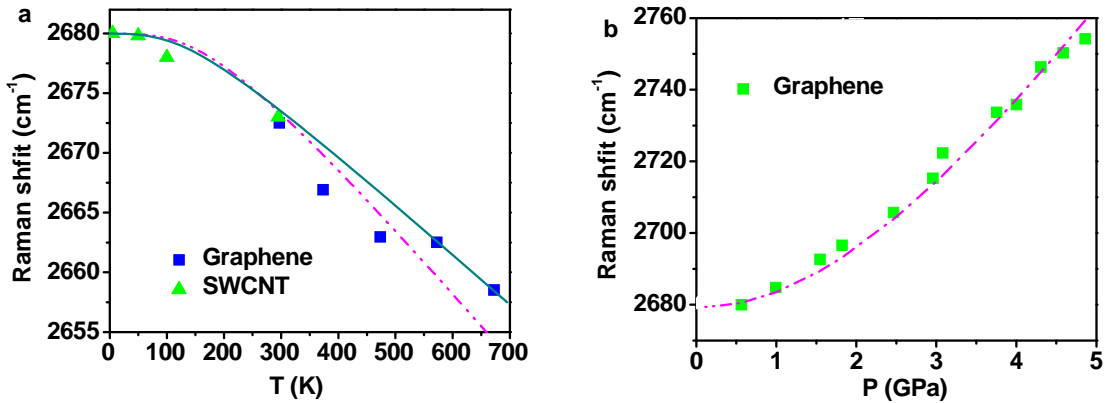

Figure 3 BOLS reproduction of the measured (a) temperature [56] and (b) pressure [8] dependent Raman shift of the 2D mode of graphene and CNT gives rise to the Debye temperature of 504 K

with the given atomic cohesive energy of 3.11 eV/atom, and the compressibility and energy density as given in Table 1.

**Table 1 Instrumental information derived from the reproduction of the ε-, T-, and P-dependent Raman shift of monolayer graphene.**

| Stimuli | Quantities | Values | Refs |
|---|---|---|---|
| Number-of-layer | $\omega_x(1)$ | $\begin{cases} 1276.8 & (D) \\ 2562.6 & (2D) \\ 1566.7 & (G) \end{cases}$ (cm$^{-1}$) | - |
| | $\omega(z_g) - \omega(1)$ | $\begin{cases} 90.2 & (D) \\ 157.4 & (2D) \\ 16.0 & (G) \end{cases}$ (cm$^{-1}$) | - |
| Strain | $\kappa$ (Nm$^{-1}$) | 6.283 | - |
| Temperature | $E_{coh}$ (eV/atom) | 3.11 | - |
| | $\theta_D$ (K) | 540 | $T_m$ = 1605[30] |
| | $\alpha$ (10$^{-6}$K$^{-1}$) | 9.0 | - |
| Pressure | $E_{den}$ (eV/nm$^3$) | 320 | - |
| | $\beta/\beta'$ (10$^{-3}$GPa$^{-1}$/GPa$^{-2}$) | 1.145/0.0763 | - |
| | $B_0/B'_0$ (GPa/-) | 690/5 | 704/1[58] |

4.4 Edge discriminative Raman reflectivity

Recent progress [29] confirmed that the Dirac-Fermi polarons preferentially generate at the zigzag-edge of graphene[59] and at graphite surface atomic vacancies[60] because of the longer and uniform √3d distance between the dangling bond electrons along the edge. As illustrated in Figure 4, the dangling bond electrons at the edges of the armchair- and the reconstructed zigzag graphene (5 and 7 atomic rings) tend to form quasi-triple-bond between the edge atoms of shorter *d* distance. The isolation and polarization of the unpaired dangling bond electrons at the zigzag edge by the locally and deeply entrapped core and bonding electrons may scatter the incident radiation substantially and hence lowers the Raman reflectivity of the D band at the zigzag edge compared with that at the arm-chaired edge.[18, 19]

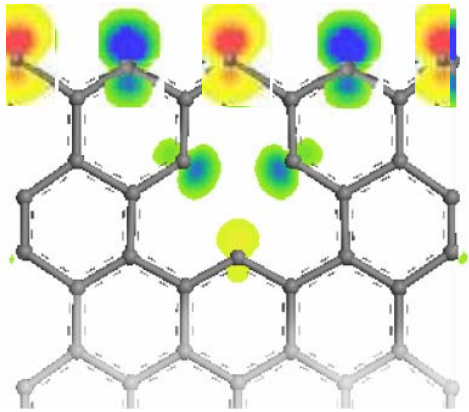

Figure 4 Density function theory calculated preferential generation of the DFPs with alternative directions (different colors) of spins at the zigzag edge and atomic vacancy of graphene. The DFPs are recognized as the isolation and polarization of the dangling bond electrons by the densely entrapped bonding electrons of the undercoordinated edge atoms that are separated by $\sqrt{3}d$ along the zigzag edge uniformly. The shorter d separation of the dangling bond electrons form quasi-triple-bond without being polarized along the arm-chair edge.[29]

5. **Conclusion**

We have formulated, for the first time, the number-of-layer, uni-axial-strain, pressure, and temperature dependent Raman shifts of graphene, as a union, as a function depending on the BOLS correlation in terms of the response of the length and energy of the representative bond to the applied stimuli. Numerical reproduction of the measurements clarified that: (i) the number-of-layer induced D and 2D redshift and the G mode blueshift are ruled by different mechanisms resulting from the undercoordination induced bond strain and bond strength gain; The D/2D vibration involves all the z neighbors on a C atom while the G mode involve only a dimer; (ii) the strain-induced phonon softening and band splitting arise from the anisotropic bond elongation and bond weakening; (iii) the thermally-softening originates from bond expansion and bond weakening due to vibration; and (iv) the mechanically-stiffening results from bond compression and bond strengthening due to mechanical work hardening. Reproduction of the measurements has led to quantitative information, as summarized in Table 1, of the referential frequencies for each mode from which the Raman shifts proceed, the bond length, bond energy, atomic cohesive energy, binding energy density, force constant, Debye temperature, compressibility, elastic modulus of graphene, and the effective coordination numbers for the few-layer graphene, which is of instrumental importance to the understanding of the unusual behavior of graphene. Findings and understandings demonstrate the essentiality of the proposed approach that has empowered the Raman spectroscopy immensely in gaining quantitative information about the dynamics of the representative bond of a specimen.

Acknowledgements

Financial supports from NSF (Nos. 50832001 and 11172254) of China and MOE (RG15/09), Singapore, are gratefully acknowledged.